\documentclass{aastex631}

\begin{document}

\title{Discovery of An X-ray Photoionized Optical Nebula and A Radio Nebula Associated with the ULX NGC 4861 X-1}

\correspondingauthor{Hang Gong, Ryan Urquhart, Jifeng Liu}
\email{ghang.naoc@gmail.com, urquha20@msu.edu, jfliu@nao.cas.cn}

\author{Hang Gong}
\affiliation{Key Laboratory of Optical Astronomy, National Astronomical Observatories, Chinese Academy of Sciences, Beijing 100101, China}

\author{Ryan Urquhart}
\affiliation{Center for Data Intensive and Time Domain Astronomy, Department of Physics and Astronomy, Michigan State University, East Lansing, MI 48824, USA}

\author{Alexandr Vinokurov}
\affil{Special Astrophysical Observatory, Nizhnij Arkhyz 369167, Russia}

\author{Yu Bai}
\affiliation{Key Laboratory of Optical Astronomy, National Astronomical Observatories, Chinese Academy of Sciences, Beijing 100101, China}
\affiliation{Institute for Frontiers in Astronomy and Astrophysics, Beijing Normal University, Beijing 102206, China}

\author{Antonio Cabrera-Lavers}
\affil{GRANTECAN, Cuesta de San Jos\'e s/n, E-38712, Bre\~na Baja, La Palma, Spain}
\affil{Instituto de Astrof\'\i sica de Canarias, V\'\i a L\'actea s/n, E-38200 La Laguna, Tenerife, Spain}

\author{Sergei Fabrika}
\affil{Special Astrophysical Observatory, Nizhnij Arkhyz 369167, Russia}

\author{Liang Wang}
\affil{Nanjing Institute of Astronomical Optics \& Technology, Chinese Academy of Sciences, Nanjing 210042, China}
\affil{CAS Key Laboratory of Astronomical Optics \& Technology, Nanjing Institute of Astronomical Optics \& Technology, Chinese Academy of Sciences, Nanjing 210042, China}

\author{Jifeng Liu}
\affiliation{Key Laboratory of Optical Astronomy, National Astronomical Observatories, Chinese Academy of Sciences, Beijing 100101, China}
\affiliation{Institute for Frontiers in Astronomy and Astrophysics, Beijing Normal University, Beijing 102206, China}
\affil{School of Astronomy and Space Sciences, University of the Chinese Academy of Sciences, Beijing 100049, China}
\affil{WHU-NAOC Joint Center for Astronomy, Wuhan University, Wuhan, Hubei 430072, China}



\begin{abstract}
We have conducted long-slit spectroscopic observations and analyzed archival radio data for the ultraluminous X-ray source (ULX) NGC 4861 X-1. Our spectral line analysis unveils that NGC 4861 X-1 is the fourth ULX situated within an X-ray photoionized nebula, following three previous findings made approximately two decades ago. Remarkably, we discover NGC 4861 X-1 also possesses a radio nebula emitting optically thin synchrotron radiation, which contradicts its X-ray photoionization and raises doubts about the four ULXs being a mere coincidence. Instead of gradually accumulating from different bands bit by bit, our multi-band discovery is made all at once.
Moreover, we tentatively perceive a faint continuum spectrum of the optical nebula. Further observations are needed to ascertain its radio structures and verify the optical continuum.

\end{abstract}



\section{Introduction} \label{sec:intro}
Ultra-luminous X-ray sources \citep[ULXs,][]{2017ARA&A..55..303K,2021AstBu..76....6F,2023NewAR..9601672K,2023arXiv230200006P} are off-nuclear X-ray sources with X-ray luminosities exceeding $1\times 10^{39}$\,$\rm{erg\,s^{-1}}$ in nearby galaxies. They were once considered to be promising intermediate-mass black hole (IMBH) candidates due to their extreme $L_{\mathrm{X}}$. However, as more and more ULX pulsars are discovered \citep[e.g.,][]{2014Natur.514..202B}, luminosity-based identifications have become feeble. It is now clear that the majority of ULXs are stellar-mass ($\leq10\,\rm M_{\odot}$) objects accreting at super-Eddington rates rather than sub-Eddington IMBHs. In the ULX family, very few \citep[e.g.,][]{2009Natur.460...73F,2014Natur.513...74P,2015MNRAS.448.1893M} are still considered to be IMBHs of decent quality.

Due to their large distances, most of the ULXs do not show optical counterparts even in $\textit{HST}$ images \citep{2013ApJS..206...14G}. Except for few ULXs \citep[e.g.,][]{2014Natur.514..198M,2015Natur.528..108L} which have optical counterparts with confirmed stellar types, most of the optically visible ULXs are buried in extended nebular powered by shocks or X-ray photoionization, or usually both \citep{2006IAUS..230..293P,2007AstBu..62...36A}. Differentiating these mechanisms requires emission line diagnostics \citep{2006agna.book.....O}, commonly achieved by the examination of line ratios encompassing elements such as hydrogen, oxygen, sulphur and nitrogen. Furthermore, He II$\lambda$4686, which needs strong UV or X-ray photons ($>$54\,eV) to produce, favors X-ray photoionization. The presence of emission line broadening implies the existence of shocks.
For instance, in the large optical bubble (452 x 266\,pc) of NGC 1313 X-1  \citep{2022A&A...666A.100G}, shock ionization was revealed by [O I]$\lambda$6300 and [SII]$\lambda$6716 in the outer regions, while a highly ionized zone (140\,pc) traced by [O III]$\lambda$5007/H$_\beta$ was detected in the interior region.

The interactions of ULXs with their surrounding environment occur across multi-bands, primarily via winds or jets, releasing vast amounts of radiation energy and mechanical power \citep{2023arXiv230200006P}. However, similarly, it is noteworthy that only a limited number of ULXs have been detected in the radio bands, indicating a relatively low probability of detection \citep{2013MNRAS.436.3128M}. A handful of their counterparts are identified to be radio bubbles \citep[e.g.,][]{2020ApJ...896..117B} inflated by jets or winds. Approximately five ULXs display radio emission formed by core or lobes. To date, Ho II X-1 \citep{2014MNRAS.439L...1C} is unique as it is the only one where the radio core and two radio lobes have been resolved. In the case of NGC 7793 S26 \citep{2010MNRAS.409..541S} and NGC\,2276\,3c \citep{2013MNRAS.436.3128M}, they each possess two radio lobes. However, M83\,MQ1 \citep{2014Sci...343.1330S} and M51\,ULX-1 \citep{2018MNRAS.475.3561U} only have a one-sided lobe.

Based on emission line diagnostics, there have been three ULXs identified to have X-ray photoionized optical nebulae. They are NGC 6946 ULX-1 \citep{1994ApJ...424L.103B,2008arXiv0809.0409A}, NGC 5408 X-1 \citep{2009ApJ...697..950K} and the aforementioned Ho II X-1 \citep{2002astro.ph..2488P,2004MNRAS.351L..83K}.
Meanwhile, they all have extended but small (20--50\,pc) radio nebulae with optically thin synchrotron emission and 5\,GHz luminosities of 2--11$\times 10^{34}$\,$\rm{erg\,s^{-1}}$ \citep{1994ApJ...425L..77V,2007ApJ...666...79L,2014MNRAS.439L...1C}, which apparently contradict their X-ray photoionization. Here we present multi-wavelength observations of the ULX NGC 4861 X-1. Through long-slit optical observations and C-band Very Large Array (VLA) archival data analysis, we demonstrate that the source is embedded in an X-ray photoionized nebula and extended radio emission. We also perceive a possible continuum spectrum of its optical nebula.

\section{The ULX and Its Data Analysis}
NGC 4861 X-1 (aka. CXOU J125901.8$+$345115), also dubbed as IXO 73 in \citet{2002ApJS..143...25C}, is a ULX located in a cometary blue compact dwarf (BCD) galaxy NGC 4861 (aka. Arp 266 or I Zw 49). BCD galaxies are characterized by their irregular morphology, with a bright off-centered star burst region that resembles a comet's head, which means Mrk 59 for NGC 4861 \citep{1967Afz.....3...55M}, and a fainter body that resembles a comet's tail. Accordingly, NGC 4861 (Fig.1, \textit{left}) contains a large number of young, massive and blue stars, resulting in strong UV emission \citep{2014MNRAS.441.1841T}. A recent IFU observation performed by the Calar Alto 3.5m telescope shows He II$\lambda$4686 emission in the central region of Mrk 59 comparable to the amount of $\approx300$ Wolf-Rayet stars \citep{2023MNRAS.523..270R}. No Cepheid variable stars have been reported in this galaxy to determine its distance.
Two slightly different distances 10.7\,Mpc \citep{2014MNRAS.441.1841T} and 9.95\,Mpc \citep{2021MNRAS.505..771O}, which are based on HI Kinematics and Tully-Fisher relation, respectively, were used in previous studies.
We simply adopt a distance of 10\,Mpc in this paper.

The main effective studies of NGC 4861 X-1 \citep{2014MNRAS.441.1841T, 2021MNRAS.505..771O} were based on $\textit{HST}$ photometry and X-ray analysis of \textit{XMM-Newton} and \textit{Chandra} data. The ULX has a persistent $L_{\mathrm{X}}$ on the level of several $10^{39}$ erg s$^{-1}$. However, despite the sparse sampling of X-ray observations (e.g., 2003-06, 2003-07 and 2003-12 by \textit{XMM-Newton}; 2012-01 and 2018-03 by \textit{Chandra}), it should be noted that NGC 4861 X-1 may vary, at least on a relatively long time scale. For example, regardless of the spectral model adopted \citep[Table.7 of][]{2021MNRAS.505..771O}, its $\textit{Chandra}$ $L_{\mathrm{X}}$ fell by a factor of at least three between the observation in January 2012 (ObsID=12473) and four observations in March 2018 (e.g., ObsID=20993), but still remained around $2\times 10^{39}$\,$\rm{erg\,s^{-1}}$. Additionally, \citet{2002ApJS..143...25C} reported that IXO 73's \textit{ROSAT}/HRI count rate increased by about five times within two observations separated by half a year.

The optical counterpart of NGC 4861 X-1 has been identified as a moderately isolated bright point source with a magnitude of $\approx21.7$ in the \textit{HST}/F814W band image \citep{2014MNRAS.441.1841T}. The authors also suggest that the point source corresponds to a late-type O supergiant or a hypergiant star. However, both \citet{2014MNRAS.441.1841T} and \citet{2021MNRAS.505..771O} also argue the optical counterpart coincides with an HII region based on its \textit{HST}/F658N image.
\citet{2021MNRAS.505..771O} also mentioned an optical spectroscopic observation conducted by Russia's BTA-6 telescope in March 2020, but only the value of A(V) was reported.

\subsection{Optical}
Here we present two optical spectroscopic observations of NGC 4861 X-1's optical counterpart in a time span of seven years.
\subsubsection{Gran Telescopio Canarias}
Our main observation (PI: Liu), which was simultaneous with a $\textit{Chandra}$ observation (ObsID=20992), was performed by the OSIRIS, the workhorse instrument of the 10.4\,m Gran Telescopio Canarias (GTC) of Spain in March 2018. A 1$\arcsec$ slit and a 30.2$^{\circ}$ slit angle from north to east (Fig.1, \textit{middle}) were utilized to encompass the galaxy's head area. However, subsequent data analysis revealed the head spectra are saturated due to its brightness. The target remained in the slit during the exposures, as confirmed by a slit image. A visual inspection of the optical counterpart suggests an obvious elongation, with the geometric center of the optical emission near the red X (Fig.1, \textit{middle}). In the slit direction, the optical counterpart has an extension of $\approx0.6\arcsec$ ($\approx29.1$\,pc at 10\,Mpc). Perpendicular to the slit, the extension is roughly $\approx0.3\arcsec$. The seeing was $<1\arcsec$.

We used {\sc IRAF} and followed standard procedures for the spectroscopic data reduction. Fig.2 demonstrates that both sub-spectra are primarily composed of emission lines. We used {\sc IRAF} routine \textit{splot} to determine the line properties. In the case of blended lines, such as the [S II] doublet, a Gaussian fit was utilized to deblend them. Uncertainties were obtained by calculating the mean values and standard deviations of these measurements, respectively \citep{2015ApJ...801L..28K}. Results for the line properties of NGC 4861 X-1 are compiled in Table.1. There are two narrow emission lines unidentified ($\lambda$5172\,$\rm{\AA}$ and $\lambda$7402\,$\rm{\AA}$ in Fig.2). They can not be attributed to second order contamination, as they did not appear in the blue grisms as R1000B\footnote{\url{http://www.gtc.iac.es/instruments/osiris/}}. Additionally, they can not be classified as ghost lines, as they typically only manifest under highly illuminated conditions such as arcs and flats. Furthermore, no strong sky lines appear at $\lambda$5172\,$\rm{\AA}$ or $\lambda$7402\,$\rm{\AA}$, leading us to conclude that the unidentified lines are from the ULX.

The optical counterpart has a negligible level of Galactic foreground extinction (A(V)=0.029 \footnote{\url{https://irsa.ipac.caltech.edu/applications/DUST/}}), due to its high Galactic latitude ($>$82.1$^{\circ}$). Assuming A(V)/E(B-V)=3.1 \citep{1989ApJ...345..245C}, the color excess E(B-V) is $\approx0.0094$, which is consistent with the value of E(B-V)=0.01 for NGC 4861 in Table.4 of \citet{2011AJ....141...37L}. After correcting for the foreground extinction, we also corrected for the main extinction caused by NGC 4861 itself. Assuming a Case B scenario, beginning with electron density $n_e$=100\,cm$^{-3}$ and electron temperature $T_e$=10000\,K, based on their corresponding theoretical Balmer line ratios H$_\alpha$/H$_\beta$=2.86, H$_\gamma$/H$_\beta$=0.468 and H$_\delta$/H$_\beta$=0.259 \citep{1995MNRAS.272...41S,2006agna.book.....O},
we derived E(B-V)$\approx0.117$, 0.537 and 0.716, respectively, if we adopt R(V)=4.05, which is common for star-burst galaxies \citep{2000ApJ...533..682C}.
Note that when $T_e$ is around 10000\,K, the intrinsic Balmer line ratios are not sensitive to $n_e$, even if $n_e$ changes between 10$^{2}$ and 10$^{6}\,\rm{cm^{-3}}$.

We speculate that in this optical nebula E(B-V), $T_e$ and $n_e$ may be not unified but complicated and position-dependent, as reported in galaxy observations \citep{2018ApJ...865...13T}, despite the optical nebula being a point source compared to the slit width. Inconsistencies in E(B-V) may not be rare, but are often obscured due to the lack of sufficient wavelength coverage in ULX observations to cover enough Balmer lines simultaneously. \citet{2008arXiv0809.0409A} briefly addressed nonuniform electron densities in MF16 (NGC 6946 ULX-1). As shown in Fig.3, the sulphur line ratio changes as the distance to the nebular center changes, demonstrating that $n_e$ should be different. Reminded by the statements about the  dust impact in MF16 \citep{2008arXiv0809.0409A}, we argue different dust destruction in different regions of our nebula might lead to position-dependent extinction. We also note that \citet{2023MNRAS.523..270R} used H$_\gamma$/H$_\beta$ to estimate the extinction due to possible sub-theoretical ratios \citep{2003A&A...407...75G,2017RAA....17...41G} in NGC 4861, which was not seen by \citet{2000A&A...361...33N}. As showed later, adopting which Balmer line ratio would not affect the recognition that the optical nebula of NGC 4861 X-1 is X-ray photoionized.

We naively adopted, like most of the papers, the E(B-V) based on H$_\alpha$/H$_\beta$ to correct reddening (the third column of Table.1) in this paper. [OIII]$\lambda\lambda$4959,5007 and [SII]$\lambda\lambda$6716,6731 are good diagnostic lines to derive $T_e$ and $n_e$, respectively. The [OII]$\lambda\lambda$3726,3729 doublet is also an indicator of $n_e$, but this doublet could not be resolved in the GTC spectra. Using the IRAF routine \textit{temden}, we obtained $T_e$=22488\,K and $n_e$=3017\,$\rm{cm^{-3}}$, respectively. Note that both line ratios are not sensitive to extinction correction due to their adjacent wavelengths. The electron density in this particular region is at least fifty times higher compared to other regions \citep{2023MNRAS.523..270R}, and at least five times higher compared to other ULX nebulae \citep{2008arXiv0809.0409A}. It is not inconsistent with that found in HII regions \citep{2009A&A...507.1327H}.

Based on Table.1, [O I]$\lambda$6300/H$_\alpha$=21.5\% indicates that NGC 4861 X-1 has warm and weakly ionized gas generated by X-ray photoionization \citep{2002astro.ph..2488P,2022A&A...666A.100G}. The [S II](6716+6731)/H$_\alpha$=28.4\% ratio and [O II](3727+3729)/[O III]$\lambda$5007=42.9\% ratio ($\lambda\lambda$3727,3729 are blended, but can be fitted with a single Gauss profile) also favor X-ray photoionization \citep{1993ApJ...407..564S,2002A&A...383...46M,2021MNRAS.501.1644S}. The impact of reddening on line ratios is not significant enough to alter the determination of whether the nebula is X-ray photoionized. Adopting H$_\gamma$/H$_\beta$ like \citet{2023MNRAS.523..270R} would only double the [O II](3727+3729)/[O III]$\lambda$5007 ratio, which still keeps this value within a safe range.

The immense L$_{4686}$ additionally supports X-ray photoionization. According to our GTC observation, NGC 4861 X-1 is one of the brightest HeII emitters among ULXs \citep{2007AstBu..62...36A,2014ApJ...797L...7G}.
Assuming spherical radiation, its corrected L$_{4686}$ amounts to 3.5$\times10^{36}$ erg s$^{-1}$. Nevertheless, compared with other three X-ray photoionized ULXs, NGC 6946 ULX-1 \citep[$2.0\times 10^{37}$\,$\rm{erg\,s^{-1}}$ in][]{2008arXiv0809.0409A}, NGC 5408 X-1 \citep[$9\times 10^{35}$\,$\rm{erg\,s^{-1}}$ in][]{2009ApJ...697..950K} and Ho II X-1
\citep[$2.7\times 10^{36}$\,$\rm{erg\,s^{-1}}$ in][]{2004MNRAS.351L..83K}, NGC 4861 X-1 just has a moderate L$_{4686}$. It is worth noting that among the four ULXs, NGC 4861 X-1's L$_{4861}$ is least compatible with its theoretical L$_{4861}$, as determined from  \citet{1989agna.book.....O}, primarily due to its notably higher electron density.
In addition, we also examined line broadenings in order to analyze the potential impact of shocks. As a result of line blendings, non-Gaussian or irregular profiles, and the relatively low resolution of the spectra, only a limited number of lines can be utilized for estimating the instrumental broadening. The FWHM of [O I]$\lambda$6300 ($\approx7.1$\,$\rm{\AA}$) is consistent with the FWHMs of the sky lines at $\lambda$6300\,$\rm{\AA}$ ($\approx7.1$\,$\rm{\AA}$), $\lambda$5577\,$\rm{\AA}$ ($\approx6.9$\,$\rm{\AA}$) and $\lambda$6863\,$\rm{\AA}$ ($\approx6.9$\,$\rm{\AA}$). Other sky lines include an emission close to $\lambda$5020, which should be a blended line \citep[$\lambda$5016 and $\lambda$5020,][]{1925RSPSA.108..501M}, and the blended sodium doublet. The narrow and isolated He II$\lambda$4686, [O III]$\lambda$4959 and [O III]$\lambda$5007 are not broadened significantly because they have consistent FWHMs and simple profiles. Together with [O I]$\lambda$6300 and [O I]$\lambda$6363, these five emission lines should have intrinsic widths $<1$\,$\rm{\AA}$, which can exclude shocks with V$>$100\,km/s safely.

It appears that there is an upward trend that peaks around $\lambda$4300\,$\rm{\AA}$ (Fig.2, \textit{bottom}). The trend is evident for wavelengths longer than $\lambda$4300\,$\rm{\AA}$. However, it becomes erratic due to the comparatively low CCD response in the blue end. If the trend is real, the continuum may arise from a companion star, an accretion disc, or the nebula itself \citep{2009ApJ...697..950K}. For the first scenario, based on the observed flux density (3--5)$\times 10^{-18}\rm{erg\,cm^{-2}\,s^{-1}\,{\AA}^{-1}}$ in the V band, the continuum would have an observed $\rm{V}\approx$ 22.2--22.7\,mag\footnote{\url{https://irsa.ipac.caltech.edu/data/SPITZER/docs/dataanalysistools/tools/pet/magtojy/}}. Taking E(B-V)$\approx0.117$ and A(V)=0.47\,mag in to account, at 10\,Mpc, this leads to an M(V) of -8.3\,mag to -7.8\,mag, which corresponds to an F-type supergiant star or a yellow hypergiant star \citep{1998A&ARv...8..145D}.

\subsubsection{BTA-6}
Prior to the BTA-6 observation performed in the blue band in March 2020 \citep{2021MNRAS.505..771O}, the other NGC 4861 X-1's BTA-6 observation was carried out using the spectrograph SCORPIO in the red band (Fig.1, \textit{right}) in February 2011. The exposure time and slit width were the same with the GTC observation. The seeing was $<1.4\arcsec$.

The types of emission lines within the common wavelength range of the red band and the GTC observations are consistent (Table.1). Nevertheless, the red band spectrum taken seven years earlier has a higher flux level, which occurred around
the same time that $\textit{Chandra}$ observed NGC 4861 in 2012. Thus, a potential correlation between the optical and X-ray bands can be established, where a higher $L_{\mathrm{X}}$ \citep[$>8\times 10^{39}$\,$\rm{erg\,s^{-1}}$ in][]{2021MNRAS.505..771O} associated with a higher optical flux level of emission lines and a lower $L_{\mathrm{X}}$ \citep[$\approx2\times 10^{39}$\,$\rm{erg\,s^{-1}}$ in][]{2021MNRAS.505..771O} corresponds to a lower optical flux level of emission lines. Nevertheless, the primary issue is that the observation conditions, especially seeing conditions, vary.
NGC 5408 X-1 \citep{2009ApJ...697..950K} was initially asserted to demonstrate temporal flux variabilities in both its optical continuum and emission lines. However, further analysis disproved this assertion and ascribed the variabilities to changes in the seeing \citep{2011ApJ...728L...5C}. Contrary to NGC 5408 X-1, for NGC 4861 X-1, the worse seeing ($<1.4\arcsec$) corresponds to a higher flux level (Fig.2, \textit{bottom}). Nevertheless, due to the significant time span between the two observations, we can only provide a preliminary indication of a correlation in flux, and it is imperative that this be further verified through more meticulous observations. The existence of the continuum of NGC 4861 X-1 is relatively reliable, but still requires verification too.

The line ratios of the two observations are consistent (Table.1). For instance, [O I]$\lambda$6300/H$_\alpha$=20.9\% and [S II](6716+6731)/H$_\alpha$=28.8\% based on the red band spectrum indicates that an X-ray photoionization scenario is still favored.
The line width analysis (Table.1) also supports that the optical nebula is X-ray photoionized.

\begin{figure}
\includegraphics[width=0.33\textwidth]{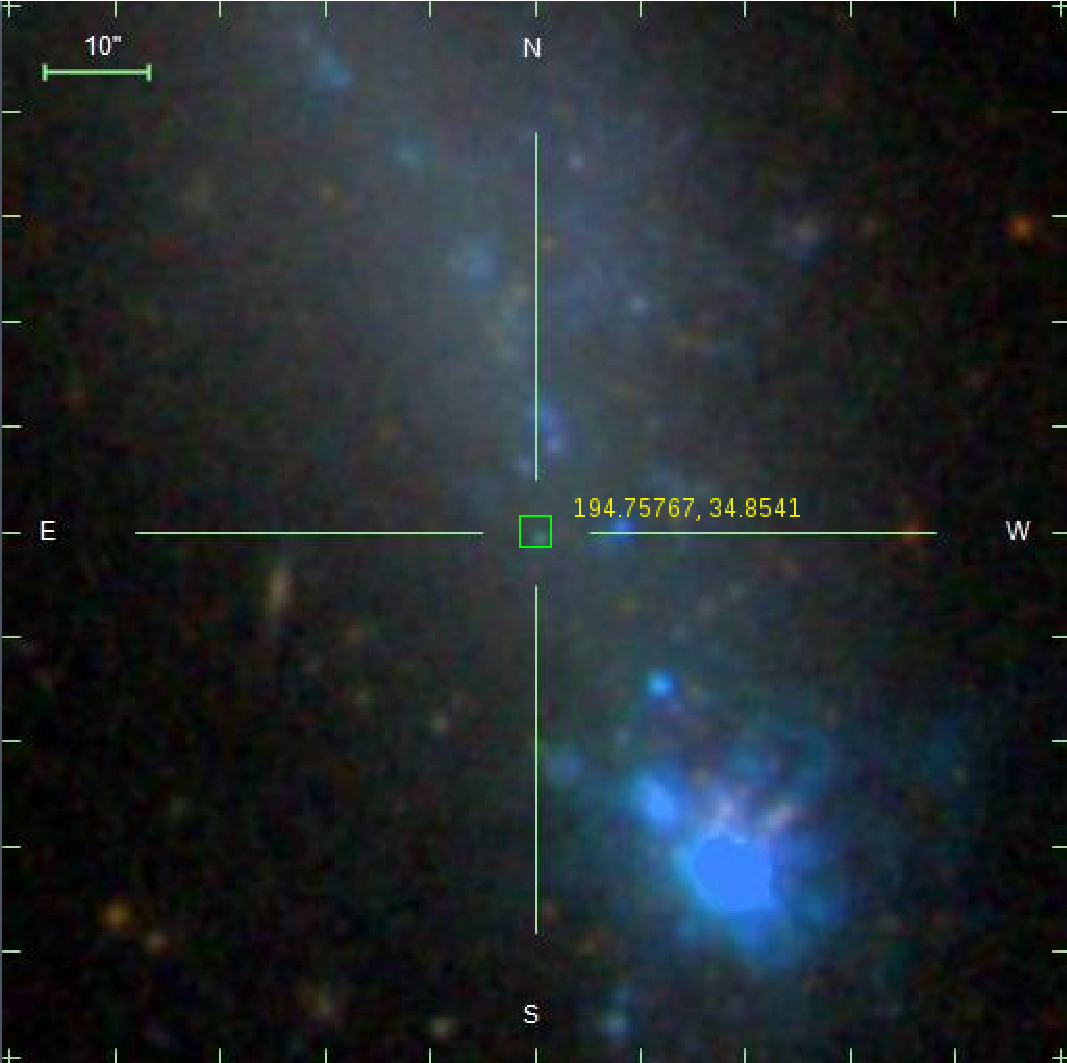}
\includegraphics[width=0.33\textwidth]{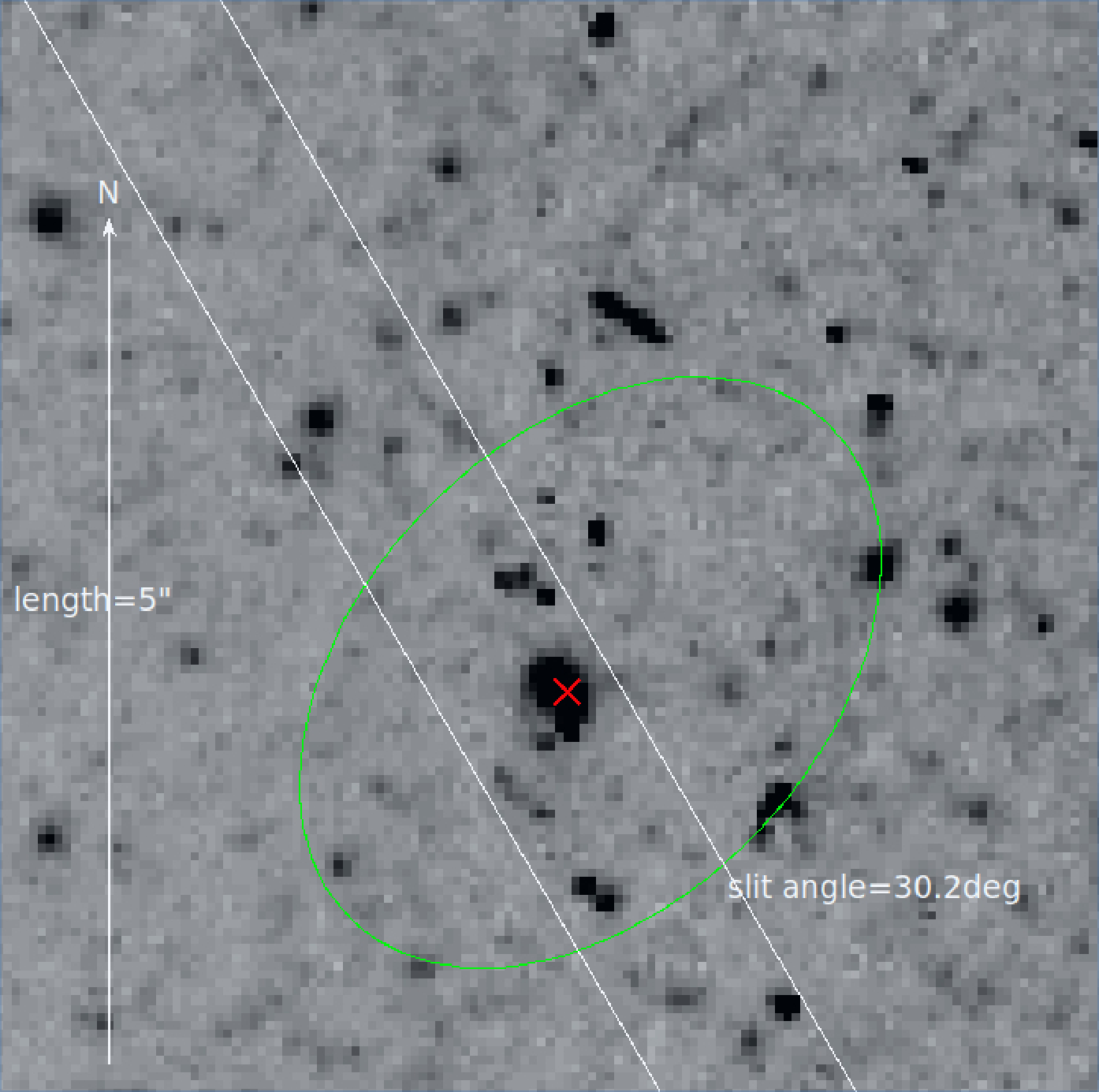}
\includegraphics[width=0.33\textwidth]{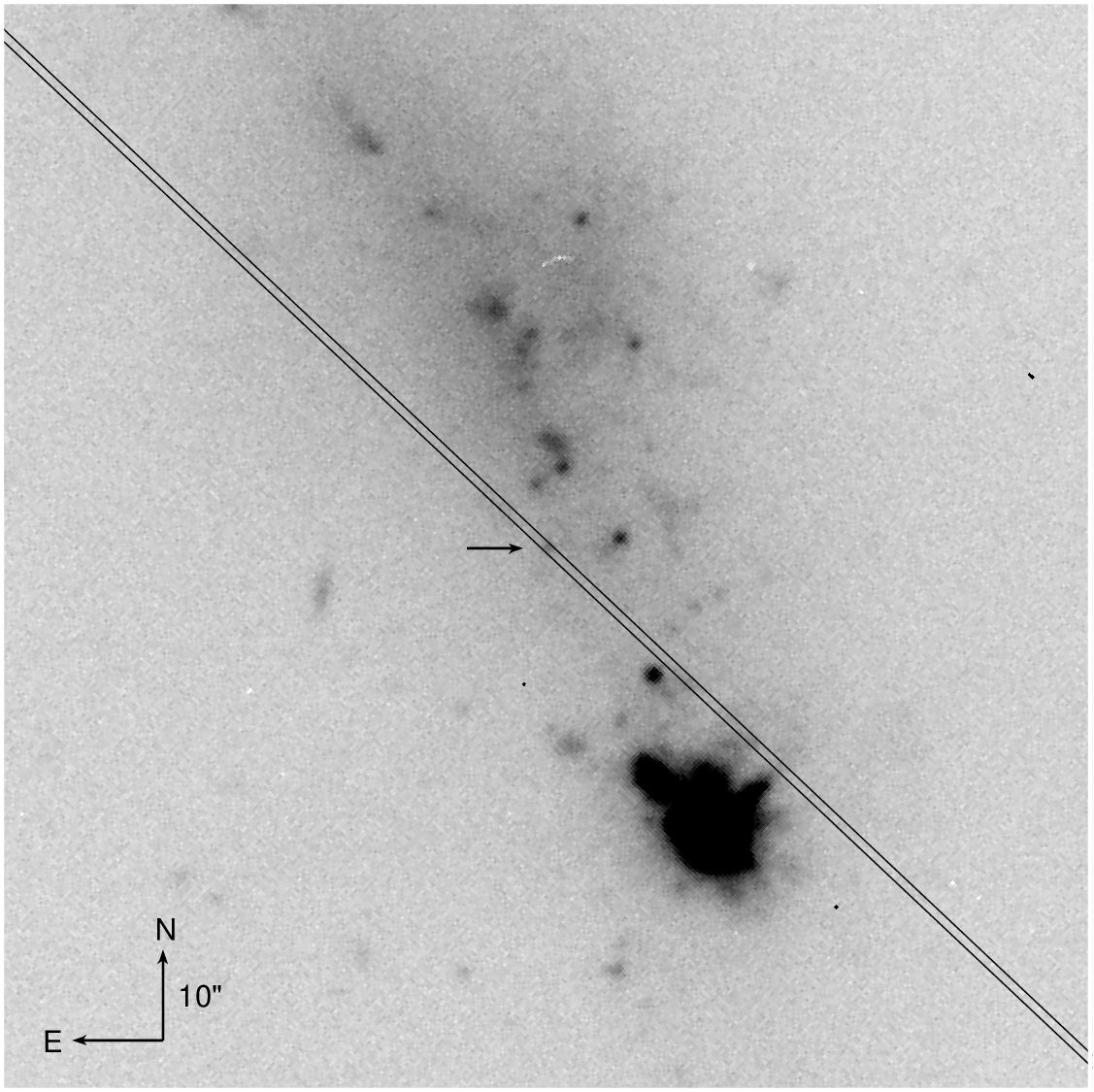}
\caption{Left panel: The SDSS image shows that NGC 4861 X-1 is located in the tail of the cometary BCD galaxy. Middle panel: The optical counterpart of NGC 4861 X-1 has an elongated structure of 0.3$\arcsec\times0.6\arcsec$ as seen in the $\textit{HST}$/F814W image. The green ellipse marks \textit{Chandra}'s 3$\sigma$ source extraction region ($r_{a}\times r_{b}=2.0\arcsec\times1.4\arcsec$) given by \textit{wavdetect} of {\sc CIAO} (ObsID=12473). The white inclined lines are the 1$\arcsec$ slit of the GTC/OSIRIS/R1000B observation. Note that NGC 4861 X-1 is not one of the two small and bright nebulae hs2 marked in Fig.2 of \citet{2023MNRAS.523..270R}, but it is clearly visible in Fig.2 and positioned slightly below and to the left of hs2. Right panel: The finding chart and 1$\arcsec$ slit of the BTA-6/SCORPIO/1200R observation in the red band. Both observations consist of two 1200\,s exposures.
\label{Fig.1}}
\end{figure}

\begin{figure}
\epsscale{1.2}
\plotone{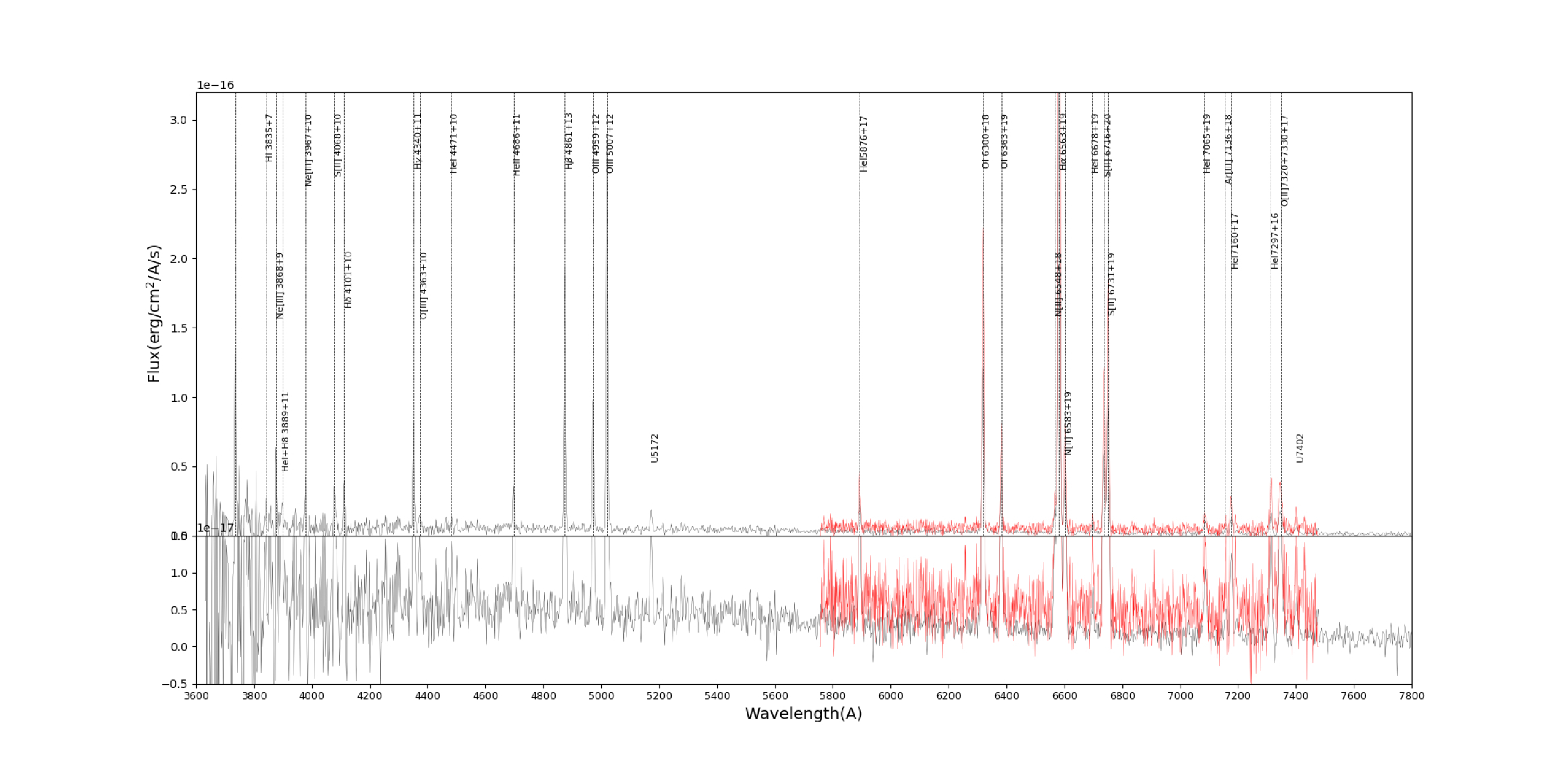}
\caption{Upper panel: GTC (black) and BTA-6 (red) spectra of NGC 4861 X-1. Lines marked with U (e.g., U5172 and U7402) are unidentified. Bottom panel: The continuum shows an upward trend and peaks at around $\lambda$4300\,$\rm{\AA}$.
\label{Fig.2}}
\end{figure}

\begin{table}
\caption{Emission line fluxes}
\begin{center}
\begin{tabular}{llllll}
\hline\hline
Lines($\rm{\AA}$) &GTC(100$\times\frac{f_{\lambda}}{f_{4861}}$)&Dereddened(H$_\alpha$/H$_\beta$)&gfwhm($\rm{\AA}$)&BTA-6(observed)    \\
\hline

[O II]$\lambda\lambda$3727,3729&60.6$\pm$0.9&118.8$\pm$1.8&5.4&\\

[Ne III]$\lambda$3868&14.6$\pm$0.4&28.0$\pm$0.8&3.9&&\\

[Ne III]$\lambda$3967&16.8$\pm$0.5&31.9$\pm$1.0&5.3&&\\

[S II]$\lambda\lambda$4067,4076&18.5$\pm$1.3&34.5$\pm$2.4&7.6&&\\

H$_\delta$$\lambda$4101&15.2$\pm$0.8&28.4$\pm$1.5&4.1&&\\

H$_\gamma$$\lambda$4340&35.9$\pm$1.0&65.0$\pm$1.8&5.4&&\\

[O III]$\lambda$4363&5.3$\pm$0.2&9.5$\pm$0.4&6.1&&\\

He II$\lambda$4686&13.0$\pm$0.3&22.5$\pm$0.5&5.7&&\\

H$_\beta$$\lambda$4861&100$\pm$1.6&170.0$\pm$2.7&6.5&&\\

[O III]$\lambda$4959&45.7$\pm$1.3&76.8$\pm$2.2&5.7&&\\

[O III]$\lambda$5007&141.4$\pm$0.8&236.1$\pm$1.3&5.8&&\\

He I$\lambda$5874&9.8$\pm$0.3&15.1$\pm$0.5&6.6&21.6$\pm$1.0&\\

[O I]$\lambda$6300&71.3$\pm$0.9&107.0$\pm$1.4&7.1&148.2$\pm$1.3&\\

[O I]$\lambda$6363&24.1$\pm$0.3&35.9$\pm$0.4&7.1&49.1$\pm$1.1&\\

[N II]$\lambda$6548&9.7$\pm$1.3&14.2$\pm$1.9&9.4&11.1$\pm$1.6&\\

H$_\alpha\lambda$6563&331.2$\pm$0.5&483.6$\pm$0.7&7.5&707.8$\pm$1.2&\\

[N II]$\lambda$6583&22.5$\pm$1.3&32.9$\pm$1.9&7.9&44.2$\pm$2.1&\\

[S II]$\lambda$6716&38.6$\pm$0.2&56.0$\pm$0.3&7.9&79.5$\pm$0.9&\\

[S II]$\lambda$6731&55.3$\pm$0.4&79.6$\pm$0.6&7.8&124.4$\pm$0.5&\\

[O II]$\lambda\lambda$7320,7330&19.6$\pm$0.7&27.4$\pm$1.0&13.8&38.5$\pm$1.7&\\
\hline
\end{tabular}

\end{center}
The dereddened fluxes are obtained by utilizing the E(B-V) determined according to the theoretical H$_\alpha$/H$_\beta$ value.
The uncorrected f$_{4861}$ and L$_{4861}$ of the GTC is $1.3\pm0.02\times 10^{-15}\rm{erg\,cm^{-2}\,s^{-1} }$ and $1.5\pm0.02\times 10^{37}\rm{erg\,s^{-1} }$, respectively.
The BTA-6's flux unit is $1\times 10^{-17}\rm{erg\,cm^{-2}\,s^{-1} }$.
\end{table}

\begin{figure}
\epsscale{0.6}
\plotone{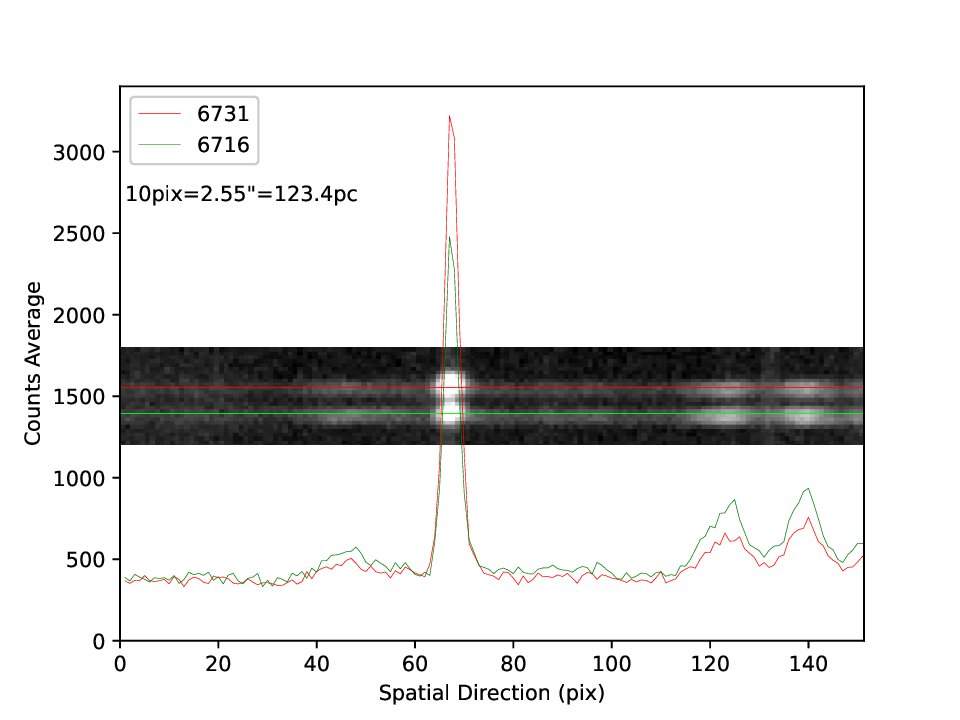}
\caption{[SII]$\lambda\lambda$6716,6731's 1D intensities based on the GTC data. Red and green horizontal lines are in the space direction. The sulfur line ratio exhibits an inverse relationship between the ULX region and its neighboring regions towards the head of the BCD.
\label{Figure 3}}
\end{figure}

\subsection{Radio} \label{sub:radio}
\subsubsection{LOFAR}
In the 2nd data release \citep{2022A&A...659A...1S} of the LOFAR Two-meter Sky Survey \citep{2017A&A...598A.104S}, we found a radio source ILTJ125901.83$+$345114.7 0.15$\arcsec$ away from the ULX. The LOFAR observation was performed one year (2019-03; field=P135$+$35\footnote{\url{https://lofar-surveys.org/observations.html}}) after the simultaneous GTC and $\textit{Chandra}$ observation. The radio source was significantly detected between 120--168\,MHz with a total flux density of 1.1$\pm$0.2\,mJy. The deconvolved major axis and minor axis of the source FWHM is 4.4$\pm1\arcsec$ and 2.1$\pm0.7\arcsec$, respectively. Considering LOFAR's 6$\arcsec$ beam size and its typical astrometric accuracy of 0.2$\arcsec$, with just a 0.15$\arcsec$ offset with the $\textit{Chandra}$ position (ObsID=12473), we argue ILTJ125901.83$+$345114.7 is the radio counterpart of NGC 4861 X-1.

\subsubsection{VLA}
NGC 4861 was observed ten times with the VLA between June and July 2015 (project code 15A-040; PI: M. Mezcua). The data were taken with C band receivers (4--8\,GHz) while the telescope was in its extended A configuration. Each observation had $\approx0.3$\,hr of on-source time, which we eventually concatenated together to achieve a total on-source time of 3\,hr. The observations only covered 4.5--6.5\,GHz of the total C band, which were then divided into two 1024-MHz-wide subbands centered at 5\,GHz and 6\,GHz (Fig.4). 3C\,286 was used as the flux and bandpass calibrator, while J1308$+$3546 was the complex gain calibrator. The data were prepared with the Common Astronomy Software Application v5.6.3 ({\sc CASA}; \citealt{2007ASPC..376..127M}). Initially, the data were processed using the {\sc CASA} (Common Astronomy Applications Software) calibration pipeline \citep{2007ASPC..376..127M}; additional flagging was performed as needed.

Coincident with the X-ray position of NGC 4861 X-1,  we have discovered a radio source (Fig.4) with peak fluxes of $74.4\pm4.1\,\mu$Jy\,beam$^{-1}$ (5\,GHz) and $61.2\pm3.5\,\mu$Jy\,beam$^{-1}$ (6\,GHz). Taking the peak fluxes results in a spectral index of -1.1$\pm$0.4 and L$_{5\,\rm{GHz}}$ =$4.5\times\rm{10^{34}}$\,erg s$^{-1}$. The radio source is mildly extended, as the integrated flux differs significantly from the peak flux. It has integrated fluxes of $117.2 \pm 6.6 \mu Jy$ (5\,GHz) and $95.2 \pm 5.4 \mu Jy$ (6\,GHz). This equates to an integrated 5\,GHz luminosity of $7\times10^{34}$\,erg s$^{-1}$.
Based on our estimation, the size of the radio counterpart is $\approx0.4\arcsec\times0.2\arcsec$ at 5\,GHz and $\approx0.3\arcsec\times0.2\arcsec$ at 6\,GHz. This suggests that the radio source could be around $\approx10$--20\,pc in size (at D=10\,Mpc). In the radio and optical bands, the extension directions of the ULX are almost consistent.

The average spectral index towards the center of the source is $\approx-0.6\pm$0.4. This steep index is consistent with optically-thin synchrotron emission, as expected for ULX bubbles or lobes. However, due to the relatively bigger errors, it is also consistent with flat ($\alpha\approx0$) bremsstrahlung emission from hot gas. Nevertheless, we can rule-out notable free-free emission from the Balmer/free-free relation \citep{1986A&A...155..297C,2010MNRAS.409..541S}. At the most, only $\approx6.7\,\mu$Jy\,beam$^{-1}$ is from free-free emission, which means the contribution of synchrotron emission dominates the observed radio luminosity. More importantly, if we consider both the LOFAR and VLA emission, the spectral index is around -0.6, which is well consistent with the in-band spectral index of the VLA.

While the astrometric accuracy of the radio source should be better than $0.1\arcsec$\footnote{\url{https://science.nrao.edu/facilities/vla/docs/manuals/oss/performance/positional-accuracy}}, in order to investigate any link between the radio source and the ULX, we need to improve the astrometry of \textit{Chandra}.
Based on five positions of archival \textit{Chandra} observations derived by \textit{wavdetect} of {\sc CIAO} 4.14, we used Equation.5 of \citet{2005ApJ...635..907H} to estimate the statistical uncertainty due to the centroiding performed by \textit{wavdetect}. We took their weighted average and weighted error, and got the final position/error at 12:59:01.823 +34:51:14.70 $\pm$0.38$\arcsec$ (the red circle in Fig.4). Alternatively, we used three \textit{GAIA} sources near the ULX to improve the \textit{Chandra} astrometry. We found in the observation (ObsID=12473) three significant X-ray sources on the same CCD of the ULX have \textit{GAIA} counterparts with a $\approx0.2''$ offset. In other four $\textit{Chandra}$ observations (e.g., ObsID=20993), although their off-axis angles are smaller, their offsets with their \textit{GAIA} counterparts are bigger (1.3$''$--1.6$''$). The bigger offsets are consistent with $\textit{Chandra}$'s astrometric accuracy degradation over time\footnote{\url{https://cxc.cfa.harvard.edu/cal/ASPECT/celmon/}}. Following the {\sc CIAO} thread of 'Correcting Absolute Astrometry' \footnote{\url{https://cxc.cfa.harvard.edu/ciao/threads/reproject_aspect/}}, we used the three \textit{GAIA} sources to correct the astrometry of $\textit{Chandra}$ observation (ObsID=12473) and obtained an updated position of the ULX at 12:59:01.798 +34:51:14.48 $\pm$0.09$\arcsec$ (the green circle in Fig.4), which is consistent with the position of the radio source. We take the \textit{GAIA}-corrected position as a better value due to the small position uncertainties ($\approx1$\,mas) of the three \textit{GAIA} sources.

\begin{figure}
\centering
\includegraphics[width=0.45\textwidth]{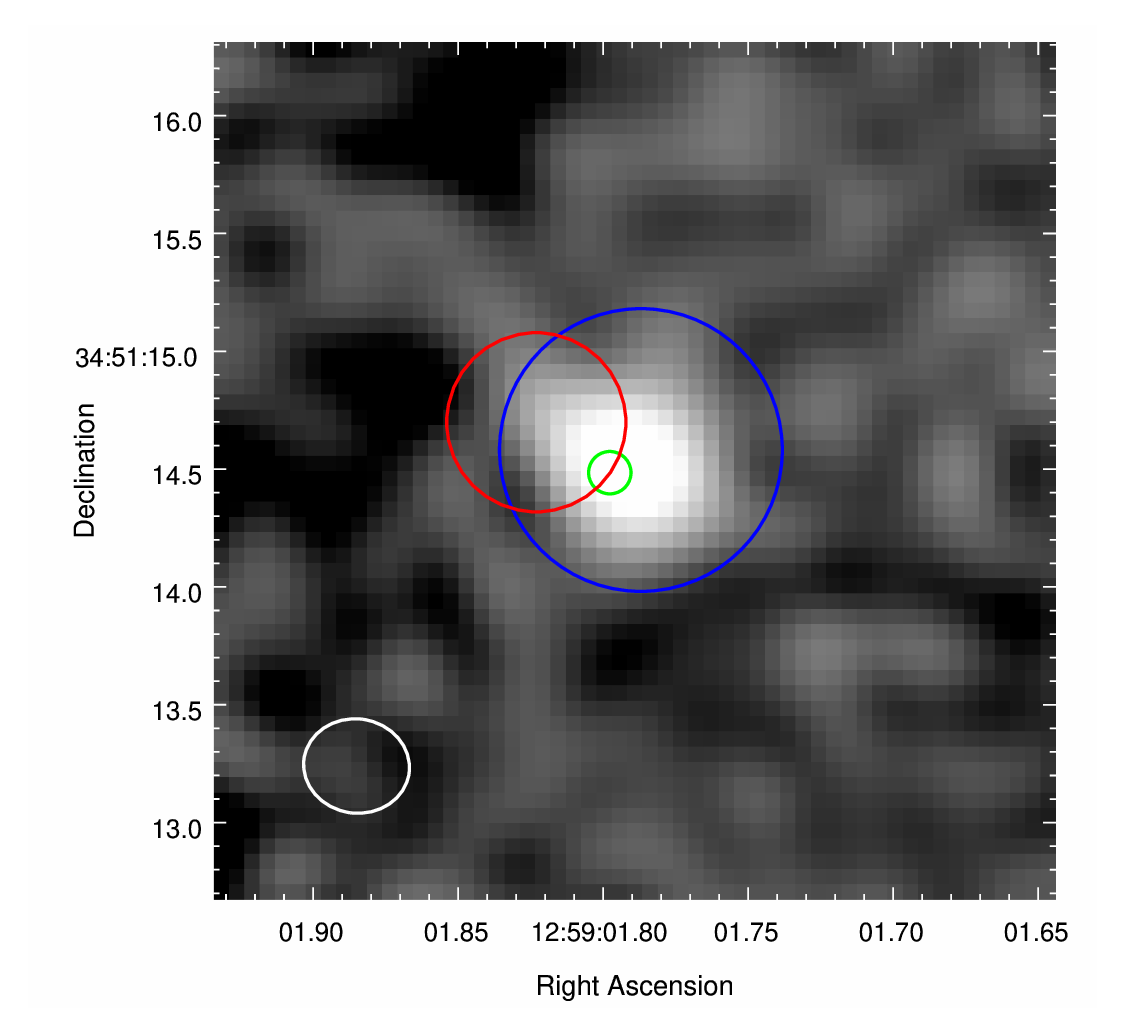}
\includegraphics[width=0.45\textwidth]{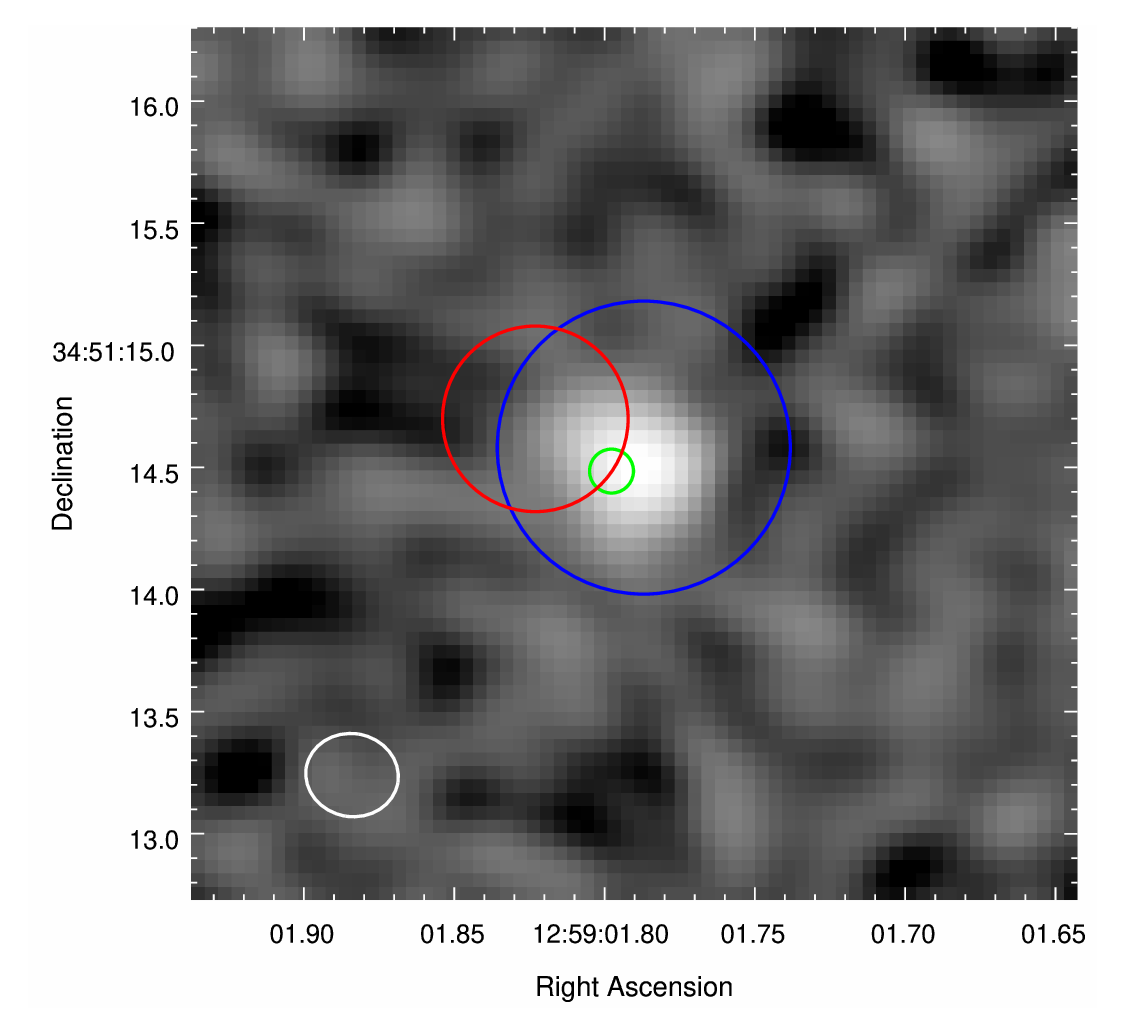}
\caption{VLA's 5\,GHz (left) and 6\,GHz (right) images in the direction of NGC 4861 X-1. The red circle (r=0.38$\arcsec$) is based on the method in \citet{2005ApJ...635..907H}. The green circle (r=0.09$\arcsec$) is based on three \textit{GAIA} sources. The blue circle, with a center of 0.16$\arcsec$ away from the center of the green circle, is without astrometric correction and derived by \textit{wavdetect} directly. Its 0.6$\arcsec$ radius is from the overall 90\% uncertainty of \textit{Chandra} X-ray absolute positions between 2007 and 2018. The synthesised beam is shown in the lower left corner of each panel.
\label{Figure 4}}
\end{figure}

\section{Discussion}
The four X-ray photoionized optical nebulae exhibit similarities in their absence of shock-ionization, substantial L$_{4686}$ and relatively moderate size.
Furthermore, despite the low probability of exhibiting strong synchrotron radio emission in the ULX family, all four ULXs demonstrate a pronounced presence of this type of radio emission. Therefore, a question arises as to whether this is merely a coincidence.
Although establishing a definitive link between X-ray photoionized nebulae and synchrotron radio emission remains challenging, this does prompt the inquiry as to why the four optical nebulae do not show any noticeable indicators of jets. In the case of Ho II X-1, the aforementioned issue can be effectively mitigated due to its comparatively smaller radius of radio influence \citep{2014MNRAS.439L...1C} compared with its extension in the optical band \citep{2002astro.ph..2488P}. The jet may be young and not long-lasting. Consequently, the interstellar medium may not have had sufficient time to enter its radiative phase. The discovery of the small triple radio structures (core+lobes) of Ho II X-1 was made only after new observations with improved angular resolution from the VLA \citep{2015MNRAS.452...24C}. As for NGC 6946 ULX-1, similarly, shock waves do not significantly contribute to the emission lines of MF16 \citep{2008arXiv0809.0409A}. The large L$_{4686}$ and L$_{4861}$ of MF16 can not be produced by its limited kinematics \citep{1996ApJS..102..161D,2008arXiv0809.0409A}. Shock waves do play a role in heating the medium and destroying dust. However, its [S II](6716+6731)/H$_\alpha$=85\% ratio, which exceeds the commonly used 40\% threshold for identifying shock-excited nebulae \citep{1994ApJ...424L.103B}, indicates that it may be in a relatively later stage prior to entering its radiative phase. Regarding NGC 5408 X-1, its synchrotron radio emission is slightly extended \citep{2007ApJ...666...79L} and relativistic X-ray outflows have been detected \citep{2016Natur.533...64P}. Additionally, its notable extension in the optical band \citep[D$\approx2.5\arcsec$ in][]{2012ApJ...745..123G} reinforced the belief held by \citet{2022A&A...666A.100G} that previous long-slit observations may have overlooked the shock-excited regions of NGC 5408 X-1. We speculate that NGC 4861 X-1 may be more like the case of Ho II X-1 due to NGC 4861 X-1's small size and small [S II](6716+6731)/H$_\alpha$ ratio. Further VLA or SKA observations with better angular resolution may help resolve the structures of NGC 4861 X-1. It would be interesting if shocks can not be detected in all regions of NGC 5408 X-1.

For the unidentified lines, despite conducting extensive searches in other ULXs that display numerous emission lines \citep[e.g.,][]{2008arXiv0809.0409A,2018MNRAS.475.3561U,2019MNRAS.489.1249L}, we have been unable to find any suitable lines in close proximity to the redshift-corrected wavelength of U5172 and U7402 (Fig.2). We note that [Fe VII]$\lambda$5159, which is in agreement with the redshift-corrected wavelength of U5172, was accompanied by a question mark among the lines of MF16 \citep{2008arXiv0809.0409A}. Besides, the progenitor of NGC 300 ULX-1 \citep{2018MNRAS.476L..45C}, SN 2010da, showed strong [Fe VII]$\lambda$5159 lines \citep{2016ApJ...830...11V}. However, their [Fe VII]$\lambda$5159 was not seen in isolation, but rather appeared alongside other iron lines, including stronger ones \citep[Fig.8 and Table.9 of][]{2016ApJ...830...11V}. Identifying U5172 encounters a similar issue when we turn to the spectra of non-ULX sources \citep[e.g.,][]{1994ApJ...435..171K,2005A&A...432..665Y,2011MNRAS.414.3360R,2015MNRAS.448.2900R}. Further observations with improved spectral resolution are necessary to help identify them.

Finally, the identification of a possible continuum and a suspicious flux correlation remains tentative. It is reasonable that a correlation is present, albeit in this specific study, it could potentially be attributed to flux calibration of the BTA-6 observation. To confirm or refute the correlation, and incidentally determine the authenticity of the continuum, repeated spectroscopic observations under different X-ray flux levels are necessary.

\begin{acknowledgments}
We would like to acknowledge Andres G\'{u}rpide, Roberto Soria, Anne Jaskot, Dmitry Bizyaev, Peter Storey, Zhou Changxing, Fang Xuan, Jin Yifei, Wang Song, Christian Motch, Manfred Pakull, Nie Jundan, Ralph Sutherland and Alberto Dominguez for their direct or indirect assistance. We appreciate the valuable suggestions provided by the reviewer.

This work is supported by the National Programs on Key Research and Development Project (Grant
No. 2019YFA0405504 and 2019YFA0405000), and the National Natural Science Foundation of China (Grant No. 11703041, 11933004 and U2031144), and the Strategic Priority Program of the Chinese Academy of Sciences (Grant No. XDB41000000).

Based on observations made with the Gran Telescopio Canarias (GTC), installed in the Spanish Observatorio del Roque de los Muchachos of the Instituto de Astrofísica de Canarias, in the island of La Palma.
Part of the observed data were obtained with a unique scientific facility, the Big Telescope Alt-azimuthal of SAO RAS; data processing and modelling of low resolution spectra were supported under the Ministry of Science and Higher
Education of the Russian Federation grant 075-15-2022-262 (13.MNPMU.21.0003).
The National Radio Astronomy Observatory is a facility of the National Science Foundation operated under cooperative agreement by Associated Universities, Inc.
This research has made use of data obtained from the $\textit{Chandra}$ Data Archive, and software provided by the $\textit{Chandra}$ X-ray Center (CXC) in the application packages {\sc CIAO}.

\end{acknowledgments}

\bibliography{sample631}{}

\begin{thebibliography}{}
\expandafter\ifx\csname natexlab\endcsname\relax\def\natexlab#1{#1}\fi
\providecommand{\url}[1]{\href{#1}{#1}}
\providecommand{\dodoi}[1]{doi:~\href{http://doi.org/#1}{\nolinkurl{#1}}}
\providecommand{\doeprint}[1]{\href{http://ascl.net/#1}{\nolinkurl{http://ascl.net/#1}}}
\providecommand{\doarXiv}[1]{\href{https://arxiv.org/abs/#1}{\nolinkurl{https://arxiv.org/abs/#1}}}

\bibitem[{{Abolmasov} {et~al.}(2007){Abolmasov}, {Fabrika}, {Sholukhova}, \&
  {Afanasiev}}]{2007AstBu..62...36A}
{Abolmasov}, P., {Fabrika}, S., {Sholukhova}, O., \& {Afanasiev}, V. 2007,
  Astrophysical Bulletin, 62, 36, \dodoi{10.1134/S199034130701004X}

\bibitem[{{Abolmasov} {et~al.}(2008){Abolmasov}, {Fabrika}, {Sholukhova}, \&
  {Kotani}}]{2008arXiv0809.0409A}
{Abolmasov}, P., {Fabrika}, S., {Sholukhova}, O., \& {Kotani}, T. 2008, arXiv
  e-prints, arXiv:0809.0409, \dodoi{10.48550/arXiv.0809.0409}

\bibitem[{{Bachetti} {et~al.}(2014){Bachetti}, {Harrison}, {Walton},
  {Grefenstette}, {Chakrabarty}, {F{\"u}rst}, {Barret}, {Beloborodov}, {Boggs},
  {Christensen}, {Craig}, {Fabian}, {Hailey}, {Hornschemeier}, {Kaspi},
  {Kulkarni}, {Maccarone}, {Miller}, {Rana}, {Stern}, {Tendulkar}, {Tomsick},
  {Webb}, \& {Zhang}}]{2014Natur.514..202B}
{Bachetti}, M., {Harrison}, F.~A., {Walton}, D.~J., {et~al.} 2014, \nat, 514,
  202, \dodoi{10.1038/nature13791}

\bibitem[{{Berghea} {et~al.}(2020){Berghea}, {Johnson}, {Secrest}, {Dudik},
  {Hennessy}, \& {El-khatib}}]{2020ApJ...896..117B}
{Berghea}, C.~T., {Johnson}, M.~C., {Secrest}, N.~J., {et~al.} 2020, \apj, 896,
  117, \dodoi{10.3847/1538-4357/ab9108}

\bibitem[{{Blair} \& {Fesen}(1994)}]{1994ApJ...424L.103B}
{Blair}, W.~P., \& {Fesen}, R.~A. 1994, \apjl, 424, L103,
  \dodoi{10.1086/187285}

\bibitem[{{Calzetti} {et~al.}(2000){Calzetti}, {Armus}, {Bohlin}, {Kinney},
  {Koornneef}, \& {Storchi-Bergmann}}]{2000ApJ...533..682C}
{Calzetti}, D., {Armus}, L., {Bohlin}, R.~C., {et~al.} 2000, \apj, 533, 682,
  \dodoi{10.1086/308692}

\bibitem[{{Caplan} \& {Deharveng}(1986)}]{1986A&A...155..297C}
{Caplan}, J., \& {Deharveng}, L. 1986, \aap, 155, 297

\bibitem[{{Cardelli} {et~al.}(1989){Cardelli}, {Clayton}, \&
  {Mathis}}]{1989ApJ...345..245C}
{Cardelli}, J.~A., {Clayton}, G.~C., \& {Mathis}, J.~S. 1989, \apj, 345, 245,
  \dodoi{10.1086/167900}

\bibitem[{{Carpano} {et~al.}(2018){Carpano}, {Haberl}, {Maitra}, \&
  {Vasilopoulos}}]{2018MNRAS.476L..45C}
{Carpano}, S., {Haberl}, F., {Maitra}, C., \& {Vasilopoulos}, G. 2018, \mnras,
  476, L45, \dodoi{10.1093/mnrasl/sly030}

\bibitem[{{Colbert} \& {Ptak}(2002)}]{2002ApJS..143...25C}
{Colbert}, E.~J.~M., \& {Ptak}, A.~F. 2002, \apjs, 143, 25,
  \dodoi{10.1086/342507}

\bibitem[{{Cseh} {et~al.}(2011){Cseh}, {Gris{\'e}}, {Corbel}, \&
  {Kaaret}}]{2011ApJ...728L...5C}
{Cseh}, D., {Gris{\'e}}, F., {Corbel}, S., \& {Kaaret}, P. 2011, \apjl, 728,
  L5, \dodoi{10.1088/2041-8205/728/1/L5}

\bibitem[{{Cseh} {et~al.}(2014){Cseh}, {Kaaret}, {Corbel}, {Grise}, {Lang},
  {Kording}, {Falcke}, {Jonker}, {Miller-Jones}, {Farrell}, {Yang}, {Paragi},
  \& {Frey}}]{2014MNRAS.439L...1C}
{Cseh}, D., {Kaaret}, P., {Corbel}, S., {et~al.} 2014, \mnras, 439, L1,
  \dodoi{10.1093/mnrasl/slt166}

\bibitem[{{Cseh} {et~al.}(2015){Cseh}, {Miller-Jones}, {Jonker}, {Gris{\'e}},
  {Paragi}, {Corbel}, {Falcke}, {Frey}, {Kaaret}, \&
  {K{\"o}rding}}]{2015MNRAS.452...24C}
{Cseh}, D., {Miller-Jones}, J.~C.~A., {Jonker}, P.~G., {et~al.} 2015, \mnras,
  452, 24, \dodoi{10.1093/mnras/stv1308}

\bibitem[{{de Jager}(1998)}]{1998A&ARv...8..145D}
{de Jager}, C. 1998, \aapr, 8, 145, \dodoi{10.1007/s001590050009}

\bibitem[{{Dopita} \& {Sutherland}(1996)}]{1996ApJS..102..161D}
{Dopita}, M.~A., \& {Sutherland}, R.~S. 1996, \apjs, 102, 161,
  \dodoi{10.1086/192255}

\bibitem[{{Fabrika} {et~al.}(2021){Fabrika}, {Atapin}, {Vinokurov}, \&
  {Sholukhova}}]{2021AstBu..76....6F}
{Fabrika}, S.~N., {Atapin}, K.~E., {Vinokurov}, A.~S., \& {Sholukhova}, O.~N.
  2021, Astrophysical Bulletin, 76, 6, \dodoi{10.1134/S1990341321010077}

\bibitem[{{Farrell} {et~al.}(2009){Farrell}, {Webb}, {Barret}, {Godet}, \&
  {Rodrigues}}]{2009Natur.460...73F}
{Farrell}, S.~A., {Webb}, N.~A., {Barret}, D., {Godet}, O., \& {Rodrigues},
  J.~M. 2009, \nat, 460, 73, \dodoi{10.1038/nature08083}

\bibitem[{{Gao} {et~al.}(2017){Gao}, {Lian}, {Kong}, {Lin}, {Hu}, {Liu},
  {Wang}, {Cao}, {Hou}, {Wang}, \& {Zhang}}]{2017RAA....17...41G}
{Gao}, Y.-L., {Lian}, J.-H., {Kong}, X., {et~al.} 2017, Research in Astronomy
  and Astrophysics, 17, 041, \dodoi{10.1088/1674-4527/17/5/41}

\bibitem[{{Gladstone} {et~al.}(2013){Gladstone}, {Copperwheat}, {Heinke},
  {Roberts}, {Cartwright}, {Levan}, \& {Goad}}]{2013ApJS..206...14G}
{Gladstone}, J.~C., {Copperwheat}, C., {Heinke}, C.~O., {et~al.} 2013, \apjs,
  206, 14, \dodoi{10.1088/0067-0049/206/2/14}

\bibitem[{{Gris{\'e}} {et~al.}(2012){Gris{\'e}}, {Kaaret}, {Corbel}, {Feng},
  {Cseh}, \& {Tao}}]{2012ApJ...745..123G}
{Gris{\'e}}, F., {Kaaret}, P., {Corbel}, S., {et~al.} 2012, \apj, 745, 123,
  \dodoi{10.1088/0004-637X/745/2/123}

\bibitem[{{G{\'u}rpide} {et~al.}(2022){G{\'u}rpide}, {Parra}, {Godet},
  {Contini}, \& {Olive}}]{2022A&A...666A.100G}
{G{\'u}rpide}, A., {Parra}, M., {Godet}, O., {Contini}, T., \& {Olive}, J.~F.
  2022, \aap, 666, A100, \dodoi{10.1051/0004-6361/202142229}

\bibitem[{{Guseva} {et~al.}(2003){Guseva}, {Papaderos}, {Izotov}, {Green},
  {Fricke}, {Thuan}, \& {Noeske}}]{2003A&A...407...75G}
{Guseva}, N.~G., {Papaderos}, P., {Izotov}, Y.~I., {et~al.} 2003, \aap, 407,
  75, \dodoi{10.1051/0004-6361:20030806}

\bibitem[{{Guti{\'e}rrez} \& {Moon}(2014)}]{2014ApJ...797L...7G}
{Guti{\'e}rrez}, C.~M., \& {Moon}, D.-S. 2014, \apjl, 797, L7,
  \dodoi{10.1088/2041-8205/797/1/L7}

\bibitem[{{Hong} {et~al.}(2005){Hong}, {van den Berg}, {Schlegel}, {Grindlay},
  {Koenig}, {Laycock}, \& {Zhao}}]{2005ApJ...635..907H}
{Hong}, J., {van den Berg}, M., {Schlegel}, E.~M., {et~al.} 2005, \apj, 635,
  907, \dodoi{10.1086/496966}

\bibitem[{{Hunt} \& {Hirashita}(2009)}]{2009A&A...507.1327H}
{Hunt}, L.~K., \& {Hirashita}, H. 2009, \aap, 507, 1327,
  \dodoi{10.1051/0004-6361/200912020}

\bibitem[{{Kaaret} \& {Corbel}(2009)}]{2009ApJ...697..950K}
{Kaaret}, P., \& {Corbel}, S. 2009, \apj, 697, 950,
  \dodoi{10.1088/0004-637X/697/1/950}

\bibitem[{{Kaaret} {et~al.}(2017){Kaaret}, {Feng}, \&
  {Roberts}}]{2017ARA&A..55..303K}
{Kaaret}, P., {Feng}, H., \& {Roberts}, T.~P. 2017, \araa, 55, 303,
  \dodoi{10.1146/annurev-astro-091916-055259}

\bibitem[{{Kaaret} {et~al.}(2004){Kaaret}, {Ward}, \&
  {Zezas}}]{2004MNRAS.351L..83K}
{Kaaret}, P., {Ward}, M.~J., \& {Zezas}, A. 2004, \mnras, 351, L83,
  \dodoi{10.1111/j.1365-2966.2004.08020.x}

\bibitem[{{Kehrig} {et~al.}(2015){Kehrig}, {V{\'\i}lchez}, {P{\'e}rez-Montero},
  {Iglesias-P{\'a}ramo}, {Brinchmann}, {Kunth}, {Durret}, \&
  {Bayo}}]{2015ApJ...801L..28K}
{Kehrig}, C., {V{\'\i}lchez}, J.~M., {P{\'e}rez-Montero}, E., {et~al.} 2015,
  \apjl, 801, L28, \dodoi{10.1088/2041-8205/801/2/L28}

\bibitem[{{King} {et~al.}(2023){King}, {Lasota}, \&
  {Middleton}}]{2023NewAR..9601672K}
{King}, A., {Lasota}, J.-P., \& {Middleton}, M. 2023, \nar, 96, 101672,
  \dodoi{10.1016/j.newar.2022.101672}

\bibitem[{{Kraemer} {et~al.}(1994){Kraemer}, {Wu}, {Crenshaw}, \&
  {Harrington}}]{1994ApJ...435..171K}
{Kraemer}, S.~B., {Wu}, C.-C., {Crenshaw}, D.~M., \& {Harrington}, J.~P. 1994,
  \apj, 435, 171, \dodoi{10.1086/174803}

\bibitem[{{Lang} {et~al.}(2007){Lang}, {Kaaret}, {Corbel}, \&
  {Mercer}}]{2007ApJ...666...79L}
{Lang}, C.~C., {Kaaret}, P., {Corbel}, S., \& {Mercer}, A. 2007, \apj, 666, 79,
  \dodoi{10.1086/519553}

\bibitem[{{Leitherer} {et~al.}(2011){Leitherer}, {Tremonti}, {Heckman}, \&
  {Calzetti}}]{2011AJ....141...37L}
{Leitherer}, C., {Tremonti}, C.~A., {Heckman}, T.~M., \& {Calzetti}, D. 2011,
  \aj, 141, 37, \dodoi{10.1088/0004-6256/141/2/37}

\bibitem[{{Liu} {et~al.}(2015){Liu}, {Bai}, {Wang}, {Justham}, {Lu}, {Gu},
  {Liu}, {di Stefano}, {Guo}, {Cabrera-Lavers}, {{\'A}lvarez}, {Cao}, \&
  {Kulkarni}}]{2015Natur.528..108L}
{Liu}, J.-F., {Bai}, Y., {Wang}, S., {et~al.} 2015, \nat, 528, 108,
  \dodoi{10.1038/nature15751}

\bibitem[{{L{\'o}pez} {et~al.}(2019){L{\'o}pez}, {Jonker}, {Heida}, {Torres},
  {Roberts}, {Walton}, {Moon}, \& {Harrison}}]{2019MNRAS.489.1249L}
{L{\'o}pez}, K.~M., {Jonker}, P.~G., {Heida}, M., {et~al.} 2019, \mnras, 489,
  1249, \dodoi{10.1093/mnras/stz2127}

\bibitem[{{Markaryan}(1967)}]{1967Afz.....3...55M}
{Markaryan}, B.~E. 1967, Astrofizika, 3, 55

\bibitem[{{McLennan} \& {Shrum}(1925)}]{1925RSPSA.108..501M}
{McLennan}, J.~C., \& {Shrum}, G.~M. 1925, Proceedings of the Royal Society of
  London Series A, 108, 501, \dodoi{10.1098/rspa.1925.0088}

\bibitem[{{McMullin} {et~al.}(2007){McMullin}, {Waters}, {Schiebel}, {Young},
  \& {Golap}}]{2007ASPC..376..127M}
{McMullin}, J.~P., {Waters}, B., {Schiebel}, D., {Young}, W., \& {Golap}, K.
  2007, in Astronomical Society of the Pacific Conference Series, Vol. 376,
  Astronomical Data Analysis Software and Systems XVI, ed. R.~A. {Shaw},
  F.~{Hill}, \& D.~J. {Bell}, 127

\bibitem[{{Mezcua} {et~al.}(2015){Mezcua}, {Roberts}, {Lobanov}, \&
  {Sutton}}]{2015MNRAS.448.1893M}
{Mezcua}, M., {Roberts}, T.~P., {Lobanov}, A.~P., \& {Sutton}, A.~D. 2015,
  \mnras, 448, 1893, \dodoi{10.1093/mnras/stv143}

\bibitem[{{Mezcua} {et~al.}(2013){Mezcua}, {Roberts}, {Sutton}, \&
  {Lobanov}}]{2013MNRAS.436.3128M}
{Mezcua}, M., {Roberts}, T.~P., {Sutton}, A.~D., \& {Lobanov}, A.~P. 2013,
  \mnras, 436, 3128, \dodoi{10.1093/mnras/stt1794}

\bibitem[{{Motch} {et~al.}(2014){Motch}, {Pakull}, {Soria}, {Gris{\'e}}, \&
  {Pietrzy{\'n}ski}}]{2014Natur.514..198M}
{Motch}, C., {Pakull}, M.~W., {Soria}, R., {Gris{\'e}}, F., \&
  {Pietrzy{\'n}ski}, G. 2014, \nat, 514, 198, \dodoi{10.1038/nature13730}

\bibitem[{{Moy} \& {Rocca-Volmerange}(2002)}]{2002A&A...383...46M}
{Moy}, E., \& {Rocca-Volmerange}, B. 2002, \aap, 383, 46,
  \dodoi{10.1051/0004-6361:20011727}

\bibitem[{{Noeske} {et~al.}(2000){Noeske}, {Guseva}, {Fricke}, {Izotov},
  {Papaderos}, \& {Thuan}}]{2000A&A...361...33N}
{Noeske}, K.~G., {Guseva}, N.~G., {Fricke}, K.~J., {et~al.} 2000, \aap, 361,
  33, \dodoi{10.48550/arXiv.astro-ph/0007130}

\bibitem[{{Osterbrock}(1989)}]{1989agna.book.....O}
{Osterbrock}, D.~E. 1989, {Astrophysics of gaseous nebulae and active galactic
  nuclei}

\bibitem[{{Osterbrock} \& {Ferland}(2006)}]{2006agna.book.....O}
{Osterbrock}, D.~E., \& {Ferland}, G.~J. 2006, {Astrophysics of gaseous nebulae
  and active galactic nuclei}

\bibitem[{{Ozdogan Ela} {et~al.}(2021){Ozdogan Ela}, {Akyuz}, {Aksaker},
  {Avdan}, {Akkaya Oralhan}, {Vinokurov}, {Allak}, {Solovyeva}, {Atapin}, \&
  {Bizyaev}}]{2021MNRAS.505..771O}
{Ozdogan Ela}, M., {Akyuz}, A., {Aksaker}, N., {et~al.} 2021, \mnras, 505, 771,
  \dodoi{10.1093/mnras/stab1321}

\bibitem[{{Pakull} {et~al.}(2006){Pakull}, {Gris{\'e}}, \&
  {Motch}}]{2006IAUS..230..293P}
{Pakull}, M.~W., {Gris{\'e}}, F., \& {Motch}, C. 2006, in Populations of High
  Energy Sources in Galaxies, ed. E.~J.~A. {Meurs} \& G.~{Fabbiano}, Vol. 230,
  293--297, \dodoi{10.1017/S1743921306008489}

\bibitem[{{Pakull} \& {Mirioni}(2002)}]{2002astro.ph..2488P}
{Pakull}, M.~W., \& {Mirioni}, L. 2002, arXiv e-prints, astro,
  \dodoi{10.48550/arXiv.astro-ph/0202488}

\bibitem[{{Pasham} {et~al.}(2014){Pasham}, {Strohmayer}, \&
  {Mushotzky}}]{2014Natur.513...74P}
{Pasham}, D.~R., {Strohmayer}, T.~E., \& {Mushotzky}, R.~F. 2014, \nat, 513,
  74, \dodoi{10.1038/nature13710}

\bibitem[{{Pinto} {et~al.}(2016){Pinto}, {Middleton}, \&
  {Fabian}}]{2016Natur.533...64P}
{Pinto}, C., {Middleton}, M.~J., \& {Fabian}, A.~C. 2016, \nat, 533, 64,
  \dodoi{10.1038/nature17417}

\bibitem[{{Pinto} \& {Walton}(2023)}]{2023arXiv230200006P}
{Pinto}, C., \& {Walton}, D.~J. 2023, arXiv e-prints, arXiv:2302.00006,
  \dodoi{10.48550/arXiv.2302.00006}

\bibitem[{{Roche} {et~al.}(2023){Roche}, {V{\'\i}lchez}, {Iglesias-P{\'a}ramo},
  {Papaderos}, {S{\'a}nchez}, {Kehrig}, \& {Duarte
  Puertas}}]{2023MNRAS.523..270R}
{Roche}, N., {V{\'\i}lchez}, J.~M., {Iglesias-P{\'a}ramo}, J., {et~al.} 2023,
  \mnras, 523, 270, \dodoi{10.1093/mnras/stad1219}

\bibitem[{{Rose} {et~al.}(2015){Rose}, {Elvis}, \&
  {Tadhunter}}]{2015MNRAS.448.2900R}
{Rose}, M., {Elvis}, M., \& {Tadhunter}, C.~N. 2015, \mnras, 448, 2900,
  \dodoi{10.1093/mnras/stv113}

\bibitem[{{Rose} {et~al.}(2011){Rose}, {Tadhunter}, {Holt}, {Ramos Almeida}, \&
  {Littlefair}}]{2011MNRAS.414.3360R}
{Rose}, M., {Tadhunter}, C.~N., {Holt}, J., {Ramos Almeida}, C., \&
  {Littlefair}, S.~P. 2011, \mnras, 414, 3360,
  \dodoi{10.1111/j.1365-2966.2011.18639.x}

\bibitem[{{Shimwell} {et~al.}(2017){Shimwell}, {R{\"o}ttgering}, {Best},
  {Williams}, {Dijkema}, {de Gasperin}, {Hardcastle}, {Heald}, {Hoang},
  {Horneffer}, {Intema}, {Mahony}, {Mandal}, {Mechev}, {Morabito}, {Oonk},
  {Rafferty}, {Retana-Montenegro}, {Sabater}, {Tasse}, {van Weeren},
  {Br{\"u}ggen}, {Brunetti}, {Chy{\.z}y}, {Conway}, {Haverkorn}, {Jackson},
  {Jarvis}, {McKean}, {Miley}, {Morganti}, {White}, {Wise}, {van Bemmel},
  {Beck}, {Brienza}, {Bonafede}, {Calistro Rivera}, {Cassano}, {Clarke},
  {Cseh}, {Deller}, {Drabent}, {van Driel}, {Engels}, {Falcke}, {Ferrari},
  {Fr{\"o}hlich}, {Garrett}, {Harwood}, {Heesen}, {Hoeft}, {Horellou},
  {Israel}, {Kapi{\'n}ska}, {Kunert-Bajraszewska}, {McKay}, {Mohan},
  {Orr{\'u}}, {Pizzo}, {Prandoni}, {Schwarz}, {Shulevski}, {Sipior}, {Smith},
  {Sridhar}, {Steinmetz}, {Stroe}, {Varenius}, {van der Werf}, {Zensus}, \&
  {Zwart}}]{2017A&A...598A.104S}
{Shimwell}, T.~W., {R{\"o}ttgering}, H.~J.~A., {Best}, P.~N., {et~al.} 2017,
  \aap, 598, A104, \dodoi{10.1051/0004-6361/201629313}

\bibitem[{{Shimwell} {et~al.}(2022){Shimwell}, {Hardcastle}, {Tasse}, {Best},
  {R{\"o}ttgering}, {Williams}, {Botteon}, {Drabent}, {Mechev}, {Shulevski},
  {van Weeren}, {Bester}, {Br{\"u}ggen}, {Brunetti}, {Callingham}, {Chy{\.z}y},
  {Conway}, {Dijkema}, {Duncan}, {de Gasperin}, {Hale}, {Haverkorn}, {Hugo},
  {Jackson}, {Mevius}, {Miley}, {Morabito}, {Morganti}, {Offringa}, {Oonk},
  {Rafferty}, {Sabater}, {Smith}, {Schwarz}, {Smirnov}, {O'Sullivan},
  {Vedantham}, {White}, {Albert}, {Alegre}, {Asabere}, {Bacon}, {Bonafede},
  {Bonnassieux}, {Brienza}, {Bilicki}, {Bonato}, {Calistro Rivera}, {Cassano},
  {Cochrane}, {Croston}, {Cuciti}, {Dallacasa}, {Danezi}, {Dettmar}, {Di
  Gennaro}, {Edler}, {En{\ss}lin}, {Emig}, {Franzen}, {Garc{\'\i}a-Vergara},
  {Grange}, {G{\"u}rkan}, {Hajduk}, {Heald}, {Heesen}, {Hoang}, {Hoeft},
  {Horellou}, {Iacobelli}, {Jamrozy}, {Jeli{\'c}}, {Kondapally}, {Kukreti},
  {Kunert-Bajraszewska}, {Magliocchetti}, {Mahatma}, {Ma{\l}ek}, {Mandal},
  {Massaro}, {Meyer-Zhao}, {Mingo}, {Mostert}, {Nair}, {Nakoneczny},
  {Nikiel-Wroczy{\'n}ski}, {Orr{\'u}}, {Pajdosz-{\'S}mierciak}, {Pasini},
  {Prandoni}, {van Piggelen}, {Rajpurohit}, {Retana-Montenegro}, {Riseley},
  {Rowlinson}, {Saxena}, {Schrijvers}, {Sweijen}, {Siewert}, {Timmerman},
  {Vaccari}, {Vink}, {West}, {Wo{\l}owska}, {Zhang}, \&
  {Zheng}}]{2022A&A...659A...1S}
{Shimwell}, T.~W., {Hardcastle}, M.~J., {Tasse}, C., {et~al.} 2022, \aap, 659,
  A1, \dodoi{10.1051/0004-6361/202142484}

\bibitem[{{Smith} {et~al.}(1993){Smith}, {Kirshner}, {Blair}, {Long}, \&
  {Winkler}}]{1993ApJ...407..564S}
{Smith}, R.~C., {Kirshner}, R.~P., {Blair}, W.~P., {Long}, K.~S., \& {Winkler},
  P.~F. 1993, \apj, 407, 564, \dodoi{10.1086/172538}

\bibitem[{{Soria} {et~al.}(2014){Soria}, {Long}, {Blair}, {Godfrey}, {Kuntz},
  {Lenc}, {Stockdale}, \& {Winkler}}]{2014Sci...343.1330S}
{Soria}, R., {Long}, K.~S., {Blair}, W.~P., {et~al.} 2014, Science, 343, 1330,
  \dodoi{10.1126/science.1248759}

\bibitem[{{Soria} {et~al.}(2010){Soria}, {Pakull}, {Broderick}, {Corbel}, \&
  {Motch}}]{2010MNRAS.409..541S}
{Soria}, R., {Pakull}, M.~W., {Broderick}, J.~W., {Corbel}, S., \& {Motch}, C.
  2010, \mnras, 409, 541, \dodoi{10.1111/j.1365-2966.2010.17360.x}

\bibitem[{{Soria} {et~al.}(2021){Soria}, {Pakull}, {Motch}, {Miller-Jones},
  {Schwope}, {Urquhart}, \& {Ryan}}]{2021MNRAS.501.1644S}
{Soria}, R., {Pakull}, M.~W., {Motch}, C., {et~al.} 2021, \mnras, 501, 1644,
  \dodoi{10.1093/mnras/staa3784}

\bibitem[{{Storey} \& {Hummer}(1995)}]{1995MNRAS.272...41S}
{Storey}, P.~J., \& {Hummer}, D.~G. 1995, \mnras, 272, 41,
  \dodoi{10.1093/mnras/272.1.41}

\bibitem[{{Thuan} {et~al.}(2014){Thuan}, {Bauer}, \&
  {Izotov}}]{2014MNRAS.441.1841T}
{Thuan}, T.~X., {Bauer}, F.~E., \& {Izotov}, Y.~I. 2014, \mnras, 441, 1841,
  \dodoi{10.1093/mnras/stu716}

\bibitem[{{Tremblay} {et~al.}(2018){Tremblay}, {Combes}, {Oonk}, {Russell},
  {McDonald}, {Gaspari}, {Husemann}, {Nulsen}, {McNamara}, {Hamer}, {O'Dea},
  {Baum}, {Davis}, {Donahue}, {Voit}, {Edge}, {Blanton}, {Bremer}, {Bulbul},
  {Clarke}, {David}, {Edwards}, {Eggerman}, {Fabian}, {Forman}, {Jones},
  {Kerman}, {Kraft}, {Li}, {Powell}, {Randall}, {Salom{\'e}}, {Simionescu},
  {Su}, {Sun}, {Urry}, {Vantyghem}, {Wilkes}, \&
  {ZuHone}}]{2018ApJ...865...13T}
{Tremblay}, G.~R., {Combes}, F., {Oonk}, J.~B.~R., {et~al.} 2018, \apj, 865,
  13, \dodoi{10.3847/1538-4357/aad6dd}

\bibitem[{{Urquhart} {et~al.}(2018){Urquhart}, {Soria}, {Johnston}, {Pakull},
  {Motch}, {Schwope}, {Miller-Jones}, \& {Anderson}}]{2018MNRAS.475.3561U}
{Urquhart}, R., {Soria}, R., {Johnston}, H.~M., {et~al.} 2018, \mnras, 475,
  3561, \dodoi{10.1093/mnras/sty014}

\bibitem[{{van Dyk} {et~al.}(1994){van Dyk}, {Sramek}, {Weiler}, {Hyman}, \&
  {Virden}}]{1994ApJ...425L..77V}
{van Dyk}, S.~D., {Sramek}, R.~A., {Weiler}, K.~W., {Hyman}, S.~D., \&
  {Virden}, R.~E. 1994, \apjl, 425, L77, \dodoi{10.1086/187314}

\bibitem[{{Villar} {et~al.}(2016){Villar}, {Berger}, {Chornock}, {Margutti},
  {Laskar}, {Brown}, {Blanchard}, {Czekala}, {Lunnan}, \&
  {Reynolds}}]{2016ApJ...830...11V}
{Villar}, V.~A., {Berger}, E., {Chornock}, R., {et~al.} 2016, \apj, 830, 11,
  \dodoi{10.3847/0004-637X/830/1/11}

\bibitem[{{Young} {et~al.}(2005){Young}, {Berrington}, \&
  {Lobel}}]{2005A&A...432..665Y}
{Young}, P.~R., {Berrington}, K.~A., \& {Lobel}, A. 2005, \aap, 432, 665,
  \dodoi{10.1051/0004-6361:20042033}

\end{thebibliography}
\bibliographystyle{aasjournal}



\end{document}